\documentclass[a4paper]{article}
\usepackage{Odyssey2024}
\usepackage{epsfig,amssymb,amsmath}
\usepackage{url}
\usepackage{multirow}
\usepackage{graphicx}
\usepackage[table, xcdraw]{xcolor}
\usepackage{xcolor}
\usepackage{tabularx}
\usepackage{subcaption}
\usepackage{comment}
\usepackage{blindtext}
\usepackage{float} 
\usepackage{array}
\usepackage{multirow}
\usepackage{ragged2e}
\usepackage{colortbl}
\usepackage{hhline}

\ninept

\title{A Comparison of Differential Performance Metrics for the \\Evaluation of Automatic Speaker Verification Fairness}

\makeatletter
\def\name#1{\gdef\@name{#1\\}}
\makeatother
\name{{\em Oubaïda Chouchane$^1$, Christoph Busch$^{2,3}$, Chiara Galdi$^1$,}\\
      {\em Nicholas Evans$^1$, Massimiliano Todisco$^1$}}

\address{$^1$EURECOM, France \\ $^2$ Hochschule Darmstadt, Germany \\ $^3$Norwegian University of Science and Technology (NTNU), Norway \\
{\small \tt$^1$\textit{firstname}[dot]\textit{lastname}[at]\textit{eurecom}[dot]\textit{fr}} \\ \small \tt$^2$\textit{firstname}[dot]\textit{lastname}[at]\textit{h-da}[dot]\textit{de}}

\begin{document}
\maketitle

\begin{abstract}
When decisions are made and when personal data is treated by automated processes, there is an expectation of fairness -- that members of different demographic groups receive equitable treatment. This expectation applies to biometric systems such as automatic speaker verification (ASV).  

We present a comparison of three candidate fairness metrics and extend previous work performed for face recognition, by examining differential performance across a range of different ASV operating points. Results show that the Gini Aggregation Rate for Biometric Equitability (GARBE) is the only one which meets three functional fairness measure criteria. 

Furthermore, a comprehensive evaluation of the fairness and verification performance of five \textit{state-of-the-art} ASV systems is also presented. Our findings reveal a nuanced trade-off between fairness and verification accuracy underscoring the complex interplay between system design, demographic inclusiveness, and verification reliability.

\end{abstract}


\section{Introduction}
The successful deployment of biometric systems is dependent upon acceptability in the target population. Acceptability requires that data subjects have confidence that they are treated in a fair manner by the biometric algorithm. The stratospheric rise in the use of artificial intelligence (AI) and its use for biometric person recognition has brought to prominence the question of whether or not it provides for the fair treatment of different demographic groups. The potential for certain demographic groups to be disadvantaged by the use of biometric person recognition systems was highlighted in the 2020 documentary film \textit{Coded Bias}\footnote{The documentary film is described at: \url{https://en.wikipedia.org/wiki/Coded_Bias}} which illustrates how a face recognition algorithm, which works well for individuals with lighter skin tone, fails to deliver the same experience to individuals with darker skin tones. Fairness is nowadays a fundamental requirement for biometric recognition algorithms.\footnote{\url{https://gdpr-info.eu/art-5-gdpr/}}

The NISTIR 8280\footnote{The draft International Standard ISO/IEC DIS 19795-10~\cite{ISO-IEC-DIS-19795-10-230928} defines differential performances as ``difference in biometric system metrics across different demographic groups".} investigated the extent of demographic-related differential performance in the case of a face recognition system. Differential performance is the difference in score distributions for mated and non-mated comparisons.\footnote{Following the International Standard ISO/IEC 19795-1~\cite{ISO-IEC-19795-1-Framework-210216} biometric performance is reported in terms of mated and non mated trials and in the form of the false match rate (FMR) and the false non-match rate (FNMR).  In the language of the speaker verification community, these are target and non-target/impostor trials, the false alarm rate and the miss rate.} 
Demographic variables can be categorical,\footnote{The draft International Standard ISO/IEC DIS 19795-10~\cite{ISO-IEC-DIS-19795-10-230928} defines categorical demographic variable as ``demographic characteristic of an individual that is nominally or ordinally described"} e.g.\ gender categories (male, female or neutral), or continuous\footnote{The draft International Standard ISO/IEC DIS 19795-10~\cite{ISO-IEC-DIS-19795-10-230928} defines continuous demographic variable as ``demographic characteristic of an individual that is observable, measurable, and that is not necessarily constrained to discrete categories"} e.g.~an individual's skin color.
There is potential for differential performance in the case of an automatic speaker verification (ASV) systems too, which can exhibit bias towards specific demographic groups, e.g.~native (or non-native) speakers of the target language. An overview of algorithmic bias in biometric systems and a survey of the recent literature can be found in~\cite{Drozdowski-BiasSurvey-TTS-2020}. The reasons for bias are manifold and range from the use of unbalanced training datasets to systematic effects in the training procedures~\cite{Drozdowski-WatchlistImbalanceEffect-ICCVW-2021, garbe, Rathgeb-FairnessExperts-TSM-2022}. 

On the path to reach fair biometric systems, a fairness testing methodology is needed. Recent proposals for fairness measures~\cite{garbe, Kotwal-FairnessIndex-ICPR-2022} are now under consideration to become the testing methodology in the draft International Standard ISO/IEC DIS 19795-10 \cite{ISO-IEC-DIS-19795-10-230928}.
To date, most of the scientific articles addressing the issue of fairness in biometric recognition systems focus on face recognition~(FR). For instance, 39 out of 54 works reviewed in~\cite{Drozdowski-BiasSurvey-TTS-2020} involve face recognition. Thus, an open question remains as to whether the criteria for selecting metrics for fairness 
assessment, such as those suggested in \cite{garbe}, retain the same characteristics when applied to different biometric traits. In this article, we present a detailed study of the most relevant metrics currently proposed for the analysis of fairness in the context of ASV.

The remainder of the paper is structured to first discuss fairness metrics and criteria in biometric recognition systems. It then presents an overview of candidate metrics: the fairness discrepancy rate~\cite{ ISO-IEC-DIS-19795-10-230928, FDR}; the inequity rate~\cite{ISO-IEC-DIS-19795-10-230928}; the Gini Aggregation Rate for Biometric Equitability~\cite{ISO-IEC-DIS-19795-10-230928,garbe}. We explain the computational methods of each, and key differences. Subsequent sections cover the experimental setup, including ASV systems, databases, and the fairness evaluation procedure. We then present experimental results and discussions, focusing on the evaluation of fairness metrics in the case of ASV systems, and conclude with a comprehensive assessment of system fairness, offering insights into performance and implications for equitable biometric recognition.

\begin{table*}[ht]
\footnotesize
\setlength{\tabcolsep}{4pt}
\caption{ASV performance in terms of pooled
EER and FMR/FNMR at the threshold corresponding to the
pooled EER, across nine groups of different nationalities. Color background transitions from better (green) to mid (yellow) to lower (red) performances.}
\label{tab:eer}
\centering
\begin{tabular}{|
>{\columncolor[HTML]{D9D9D9}}c |cccccccccc|}
\hline
\cellcolor[HTML]{D9D9D9}                                       & \multicolumn{10}{c|}{\cellcolor[HTML]{D9D9D9}\textbf{ASV Systems}}                                                                                                                                                                                                                                                                                                                                                                                                                                                                                            \\ \cline{2-11} 
\cellcolor[HTML]{D9D9D9}                                       & \multicolumn{2}{c|}{\cellcolor[HTML]{D9D9D9}\textit{\textbf{ERes2Net}}}                                        & \multicolumn{2}{c|}{\cellcolor[HTML]{D9D9D9}\textit{\textbf{CAM++}}}                                           & \multicolumn{2}{c|}{\cellcolor[HTML]{D9D9D9}\textit{\textbf{ECAPA}}}                                           & \multicolumn{2}{c|}{\cellcolor[HTML]{D9D9D9}\textit{\textbf{ResNetSE34V2}}}                                    & \multicolumn{2}{c|}{\cellcolor[HTML]{D9D9D9}\textit{\textbf{ResNetSE34L}}}                \\ \cline{2-11} 
\multirow{-3}{*}{\cellcolor[HTML]{D9D9D9}\textbf{Nationality}} & \multicolumn{1}{c|}{\cellcolor[HTML]{EFEFEF}FMR (\%)} & \multicolumn{1}{c|}{\cellcolor[HTML]{EFEFEF}FNMR (\%)} & \multicolumn{1}{c|}{\cellcolor[HTML]{EFEFEF}FMR (\%)} & \multicolumn{1}{c|}{\cellcolor[HTML]{EFEFEF}FNMR (\%)} & \multicolumn{1}{c|}{\cellcolor[HTML]{EFEFEF}FMR (\%)} & \multicolumn{1}{c|}{\cellcolor[HTML]{EFEFEF}FNMR (\%)} & \multicolumn{1}{c|}{\cellcolor[HTML]{EFEFEF}FMR (\%)} & \multicolumn{1}{c|}{\cellcolor[HTML]{EFEFEF}FNMR (\%)} & \multicolumn{1}{c|}{\cellcolor[HTML]{EFEFEF}FMR (\%)} & \cellcolor[HTML]{EFEFEF}FNMR (\%) \\ \hline
\textbf{USA}                                                   & \multicolumn{1}{c|}{\cellcolor[HTML]{F9F0CE}1.22}     & \multicolumn{1}{c|}{\cellcolor[HTML]{F4EFCE}1.04}      & \multicolumn{1}{c|}{\cellcolor[HTML]{FDF1CD}1.40}     & \multicolumn{1}{c|}{\cellcolor[HTML]{FFF2CC}1.54}      & \multicolumn{1}{c|}{\cellcolor[HTML]{F6F0CE}1.13}     & \multicolumn{1}{c|}{\cellcolor[HTML]{FFF2CC}1.45}      & \multicolumn{1}{c|}{\cellcolor[HTML]{FCF1CD}1.36}     & \multicolumn{1}{c|}{\cellcolor[HTML]{FFF1CC}1.68}      & \multicolumn{1}{c|}{\cellcolor[HTML]{FBE7C7}2.76}     & \cellcolor[HTML]{FDEDCA}2.08      \\ \hline
\textbf{UK}                                                    & \multicolumn{1}{c|}{\cellcolor[HTML]{EAEDD0}0.68}     & \multicolumn{1}{c|}{\cellcolor[HTML]{E4ECD1}0.45}      & \multicolumn{1}{c|}{\cellcolor[HTML]{E3ECD2}0.41}     & \multicolumn{1}{c|}{\cellcolor[HTML]{DCEAD3}0.14}      & \multicolumn{1}{c|}{\cellcolor[HTML]{ECEED0}0.72}     & \multicolumn{1}{c|}{\cellcolor[HTML]{DEEBD2}0.23}      & \multicolumn{1}{c|}{\cellcolor[HTML]{E6ECD1}0.50}     & \multicolumn{1}{c|}{\cellcolor[HTML]{E6ECD1}0.50}      & \multicolumn{1}{c|}{\cellcolor[HTML]{FDEDCA}2.08}     & \cellcolor[HTML]{FEEFCB}1.90      \\ \hline
\textbf{Germany}                                               & \multicolumn{1}{c|}{\cellcolor[HTML]{E8EDD1}0.59}     & \multicolumn{1}{c|}{\cellcolor[HTML]{FAE7C7}2.81}      & \multicolumn{1}{c|}{\cellcolor[HTML]{E9EDD0}0.63}     & \multicolumn{1}{c|}{\cellcolor[HTML]{F5DAC0}4.26}      & \multicolumn{1}{c|}{\cellcolor[HTML]{FCE9C8}2.49}     & \multicolumn{1}{c|}{\cellcolor[HTML]{EDC8B7}6.34}      & \multicolumn{1}{c|}{\cellcolor[HTML]{FCEBC9}2.31}     & \multicolumn{1}{c|}{\cellcolor[HTML]{EDC8B7}6.34}      & \multicolumn{1}{c|}{\cellcolor[HTML]{F5DBC1}4.17}     & \cellcolor[HTML]{E6B8AF}8.11      \\ \hline
\textbf{Australia}                                             & \multicolumn{1}{c|}{\cellcolor[HTML]{EAEDD0}0.68}     & \multicolumn{1}{c|}{\cellcolor[HTML]{E0EBD2}0.27}      & \multicolumn{1}{c|}{\cellcolor[HTML]{FDEECA}1.99}     & \multicolumn{1}{c|}{\cellcolor[HTML]{DCEAD3}0.14}      & \multicolumn{1}{c|}{\cellcolor[HTML]{FEEFCB}1.90}     & \multicolumn{1}{c|}{\cellcolor[HTML]{E0EBD2}0.27}      & \multicolumn{1}{c|}{\cellcolor[HTML]{EFEECF}0.86}     & \multicolumn{1}{c|}{\cellcolor[HTML]{EDEED0}0.77}      & \multicolumn{1}{c|}{\cellcolor[HTML]{FFF2CC}1.54}     & \cellcolor[HTML]{FEF0CB}1.72      \\ \hline
\textbf{Italy}                                                 & \multicolumn{1}{c|}{\cellcolor[HTML]{FEEECA}1.95}     & \multicolumn{1}{c|}{\cellcolor[HTML]{FBE9C8}2.58}      & \multicolumn{1}{c|}{\cellcolor[HTML]{FDF1CD}1.40}     & \multicolumn{1}{c|}{\cellcolor[HTML]{FBE7C7}2.72}      & \multicolumn{1}{c|}{\cellcolor[HTML]{FBE9C8}2.58}     & \multicolumn{1}{c|}{\cellcolor[HTML]{F6DEC2}3.85}      & \multicolumn{1}{c|}{\cellcolor[HTML]{F9E3C5}3.26}     & \multicolumn{1}{c|}{\cellcolor[HTML]{FBE9C8}2.54}      & \multicolumn{1}{c|}{\cellcolor[HTML]{F8E0C3}3.53}     & \cellcolor[HTML]{F6DCC1}4.08      \\ \hline
\textbf{India}                                                 & \multicolumn{1}{c|}{\cellcolor[HTML]{FCEBC9}2.31}     & \multicolumn{1}{c|}{\cellcolor[HTML]{DBEAD3}0.09}      & \multicolumn{1}{c|}{\cellcolor[HTML]{FEEFCB}1.90}     & \multicolumn{1}{c|}{\cellcolor[HTML]{DCEAD3}0.14}      & \multicolumn{1}{c|}{\cellcolor[HTML]{FBE7C7}2.76}     & \multicolumn{1}{c|}{\cellcolor[HTML]{F3EFCF}1.00}      & \multicolumn{1}{c|}{\cellcolor[HTML]{EECAB8}6.11}     & \multicolumn{1}{c|}{\cellcolor[HTML]{D9EAD3}0.00}      & \multicolumn{1}{c|}{\cellcolor[HTML]{F0CFBB}5.53}     & \cellcolor[HTML]{DBEAD3}0.09      \\ \hline
\textbf{Ireland}                                               & \multicolumn{1}{c|}{\cellcolor[HTML]{DDEBD3}0.18}     & \multicolumn{1}{c|}{\cellcolor[HTML]{FDEDCA}2.04}      & \multicolumn{1}{c|}{\cellcolor[HTML]{EEEED0}0.82}     & \multicolumn{1}{c|}{\cellcolor[HTML]{FCEBC9}2.31}      & \multicolumn{1}{c|}{\cellcolor[HTML]{EDEED0}0.77}     & \multicolumn{1}{c|}{\cellcolor[HTML]{FFF2CC}1.45}      & \multicolumn{1}{c|}{\cellcolor[HTML]{E4ECD1}0.45}     & \multicolumn{1}{c|}{\cellcolor[HTML]{FCEBC9}2.36}      & \multicolumn{1}{c|}{\cellcolor[HTML]{EFEECF}0.86}     & \cellcolor[HTML]{F5DAC0}4.21      \\ \hline
\textbf{New\_Zealand}                                          & \multicolumn{1}{c|}{\cellcolor[HTML]{FEEFCB}1.86}     & \multicolumn{1}{c|}{\cellcolor[HTML]{E0EBD2}0.27}      & \multicolumn{1}{c|}{\cellcolor[HTML]{FEEFCB}1.86}     & \multicolumn{1}{c|}{\cellcolor[HTML]{DDEBD3}0.18}      & \multicolumn{1}{c|}{\cellcolor[HTML]{FDECC9}2.17}     & \multicolumn{1}{c|}{\cellcolor[HTML]{E0EBD2}0.27}      & \multicolumn{1}{c|}{\cellcolor[HTML]{F1EFCF}0.95}     & \multicolumn{1}{c|}{\cellcolor[HTML]{F7F0CE}1.18}      & \multicolumn{1}{c|}{\cellcolor[HTML]{FEEFCB}1.86}     & \cellcolor[HTML]{FBE8C7}2.67      \\ \hline
\textbf{Canada}                                                & \multicolumn{1}{c|}{\cellcolor[HTML]{F6F0CE}1.13}     & \multicolumn{1}{c|}{\cellcolor[HTML]{FBF1CD}1.31}      & \multicolumn{1}{c|}{\cellcolor[HTML]{FDF1CD}1.40}     & \multicolumn{1}{c|}{\cellcolor[HTML]{EFEECF}0.86}      & \multicolumn{1}{c|}{\cellcolor[HTML]{FBF1CD}1.31}     & \multicolumn{1}{c|}{\cellcolor[HTML]{F9F0CE}1.22}      & \multicolumn{1}{c|}{\cellcolor[HTML]{F6F0CE}1.13}     & \multicolumn{1}{c|}{\cellcolor[HTML]{FFF1CC}1.63}      & \multicolumn{1}{c|}{\cellcolor[HTML]{F1D1BC}5.30}     & \cellcolor[HTML]{FDECC9}2.17      \\ \hline
\textbf{Pooled EER (\%)}                                       & \multicolumn{2}{c|}{\cellcolor[HTML]{D9EAD3}\textbf{1.18}}                                                     & \multicolumn{2}{c|}{\cellcolor[HTML]{E4ECD1}\textbf{1.37}}                                                     & \multicolumn{2}{c|}{\cellcolor[HTML]{FFF2CC}\textbf{1.79}}                                                     & \multicolumn{2}{c|}{\cellcolor[HTML]{FEEECA}\textbf{1.88}}                                                     & \multicolumn{2}{c|}{\cellcolor[HTML]{E6B8AF}\textbf{3.01}}                                \\ \hline
\end{tabular}
\end{table*}

\section{Fairness Metrics and Criteria}
\label{sec:metrics}
This section introduces fairness metrics initially proposed for the evaluation of face recognition systems, alongside essential criteria that such metrics must meet to effectively assess fairness. These guidelines ensure that the metrics are not only theoretically robust but also practically relevant in real-world applications.

\subsection{Fairness Discrepancy Rate}
\label{sec:fdr}
The fairness discrepancy rate (FDR) focuses on the balance between the false match rate and the false non-match rate in assessing demographic-related differential performance~\cite{ISO-IEC-DIS-19795-10-230928,FDR}. The measure introduces two components, namely the false positive differential (FPD) and the false negative differential (FND), which represent the maximum discrepancy in FMR and FNMR, respectively, between any two demographic groups $d_i$ and $d_j$, belonging to a set $D$, at a specific discrimination threshold $\tau$. The FDR quantifies these discrepancies, modulated by risk parameters $\alpha$ and $1 - \alpha$, to weight the relative importance of FMR and FNMR differences according to the security needs of a given application. High-risk situations, for instance, require a lower FMR to minimise security breaches. The FDR value ranges from 0 to 1, with 1 indicating full fairness and 0 indicating full unfairness. The FDR is calculated according to:

\begin{align}
\text{FPD}(\tau) &= \max \left( \left| \text{FMR}_{d_i}(\tau) - \text{FMR}_{d_j}(\tau) \right| \right) \quad  \label{eq:A_tau} \\
\text{FND}(\tau) &= \max \left( \left| \text{FNMR}_{d_i}(\tau) - \text{FNMR}_{d_j}(\tau) \right| \right) \quad \label{eq:B_tau} 
\end{align}
\noindent for any  $d_i, d_j \in D$.
\begin{align}
\text{FDR}(\tau,\alpha) &= 1 - \left( \alpha \text{FPD}(\tau) + (1-\alpha)\text{FND}(\tau) \right) \label{eq:FDR_tau}
\end{align}
In contrast to the goal in this paper, namely the comparison of different candidate metrics, the FDR metric has been used previously for the evaluation of ASV system fairness~\cite{estevez2022study,peri2023study,chouchane2023fairness}.

\subsection{Inequity Rate}
\label{sec:ir}
The inequity rate (IR) assesses fairness by examining the ratio of the boundary (minimum and maximum) FMR and FNMR values across different demographic groups $d_i$ and $d_j$~\cite{ISO-IEC-DIS-19795-10-230928}. This is accomplished by evaluating the ratio of the highest FMR and FNMR to the lowest FMR and FNMR across all groups.
Risk parameters $\alpha$ and $1 - \alpha$ are again used to scale the ratios before their aggregation. Importantly, in contrast to the the FDR, lower IR values indicate greater fairness. The IR is calculated according to:
\begin{align}
\text{FPD}(\tau) &= \frac{\max_{d_i} \text{FMR}_{d_i}(\tau)}{\min_{d_j} \text{FMR}_{d_j}(\tau)} & \forall d_i, d_j \in D \label{eq:A_IR}\\
\text{FND}(\tau) &= \frac{\max_{d_i} \text{FNMR}_{d_i}(\tau)}{\min_{d_j} \text{FNMR}_{d_j}(\tau)} & \forall d_i, d_j \in D \label{eq:B_IR}\\
\text{IR}(\tau,\alpha) &= \text{FPD}(\tau)^{\alpha} \cdot \text{FND}(\tau)^{(1-\alpha)} \label{eq:IR}
\end{align}

\subsection{The Gini Aggregation Rate for Biometric Equitability}
\label{sec:garbe}
The Gini aggregation rate for biometric equitability (GARBE) is derived from the Gini index, a measure for inequality~\cite{ISO-IEC-DIS-19795-10-230928,garbe}. The GARBE metric uses a normalised  Gini coefficient for $n$ demographic groups. The normalization by $\frac{n}{n-1}$ is proposed in~\cite{deltas2003small} to correct the downward bias when the number of samples (demographic groups) is small.

The Gini coefficient related to the FMR, is defined by:
\begin{equation}
G_{\text{FMR}}(\tau) = \frac{n}{n-1} \left( \frac{\sum_{i=1}^{n}\sum_{j=1}^{n} | \text{FMR}_{d_i}(\tau) - \text{FMR}_{d_j}(\tau)|}{2n^2\overline{\text{FMR}(\tau)}} \right)
\label{eq:Gb_FMR}
\end{equation}
where $\overline{\text{FMR}}$ is the mean value.  
Similarly, the Gini coefficient related to the FNMR is defined by:
\begin{equation}
G_{\text{FNMR}}(\tau) = \frac{n}{n-1} \left( \frac{\sum_{i=1}^{n}\sum_{j=1}^{n} | \text{FNMR}_{d_i}(\tau) - \text{FNMR}_{d_j}(\tau)|}{2n^2\overline{\text{FNMR}(\tau)}} \right)
\label{eq:Gb_FNMR}
\end{equation}

\noindent for any  $d_i, d_j \in D$.

In adapting to be consistent with the  notation above, the pair of Gini coefficients are combined according to:
\begin{equation}
\text{FPD}(\tau) = G_{\text{FMR}}, \quad \text{FND}(\tau) = G_{\text{FNMR}}
\label{eq:AB_tau}
\end{equation}
\begin{equation}
\text{GARBE}(\tau,\alpha) = \alpha \text{FPD}(\tau) + (1-\alpha) \text{FND}(\tau)
\label{eq:GARBE_tau}
\end{equation}

GARBE values range from 0 to 1, with 0 indicating full fairness and 1 indicating full unfairness.   

\subsection{Functional Fairness Measure Criteria}
\label{sec:FFMCs}
The fundamental aim of fairness metrics is to evaluate and identify the most equitable classification algorithms. Howard et al.~\cite{face_metrics_eval} define the requirement for essential attributes of such metrics as the functional fairness measure criteria (FFMC), which reflect their interpretability and applicability:

\begin{enumerate}
\item \textbf{FFMC.1}: The contributions of FMR and FNMR to the fairness metric should be intuitive across typical risk parameters and operationally relevant error rates. 
\item \textbf{FFMC.2}: The metric needs defined boundaries, with minimum and maximum values, to establish clear benchmarks.
\item \textbf{FFMC.3}: The metric must remain computable for demographic groups with no observed errors, a condition becoming more common with ever-more accurate biometric algorithms. 
\end{enumerate}

\section{Experimental setup}
In this section we present the ASV systems used for the fairness metrics evaluation, database and the fairness evaluation procedure.

 \subsection{Speaker verification systems}
We use five different ASV systems in our assessment of the fairness metrics, with each system manifesting unique structural and functional characteristics. The ECAPA system~\cite{ecapa_repo} employs a standard ECAPA-TDNN~\cite{ecapa} structure with 3 SE-Res2Block modules to extract a 192-dimensional speaker embedding. As a back-end, ECAPA uses the cosine similarity. ResNetSE34L~\cite{ResNetSE34L} is a slimmed-down version of the original ResNet-34~\cite{resnet}. It uses self-attentive pooling (SAP)~\cite{SAP} for aggregating frame-level features into utterance-level features, which focuses on the most informative frames. ResNetSE34L uses the squared Euclidean distance as a distance metric. ResNetSE34V2~\cite{ResNetSE34V2} is a performance-optimised variant of ResNet-34. The stride is removed at the first convolutional layer to reduce computational cost. ResNetSE34V2 adopts attentive statistics pooling (ASP)~\cite{ASP} for temporal frames aggregation. ERes2Net~\cite{ERes2Net} addresses the limitations of the Res2Net structure by integrating local and global feature fusion, capturing both detailed and holistic patterns in the input signal. Last, the CAM++~\cite{cam++} architecture mainly consist of a front-end convolution module and a densely connected time delay neural network (D-TDNN) backbone. In each D-TDNN layer, an improved context-aware masking (CAM) module is included. The CAM++ system employs multi-granularity pooling to capture discriminative speaker characteristics.

\begin{figure*}[t]
    \centering
    \includegraphics[width=1\textwidth]{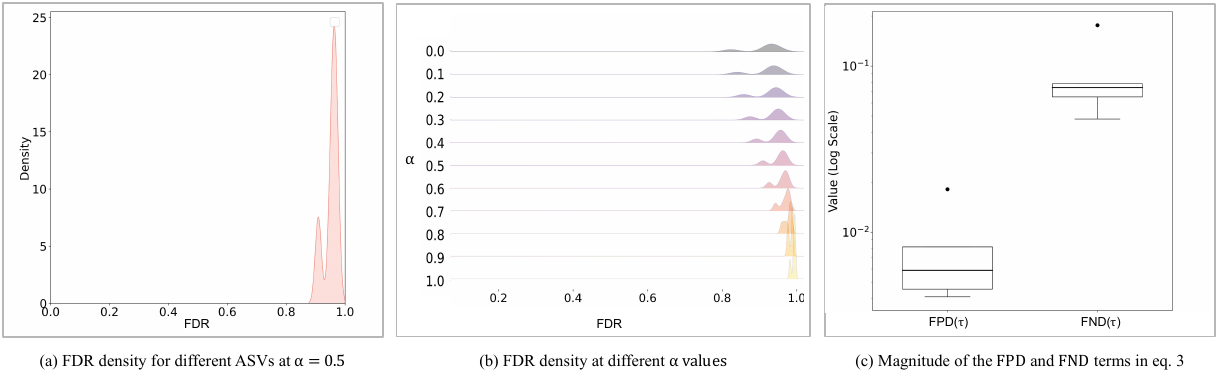}
    \caption{FDR values using 5 automatic speaker verification systems at a threshold corresponding to FMR = 0.1\%}
    \label{fig:three_figures_FDR_part}
\end{figure*}
\begin{figure*}[t]
    \centering
    \includegraphics[width=1\textwidth]{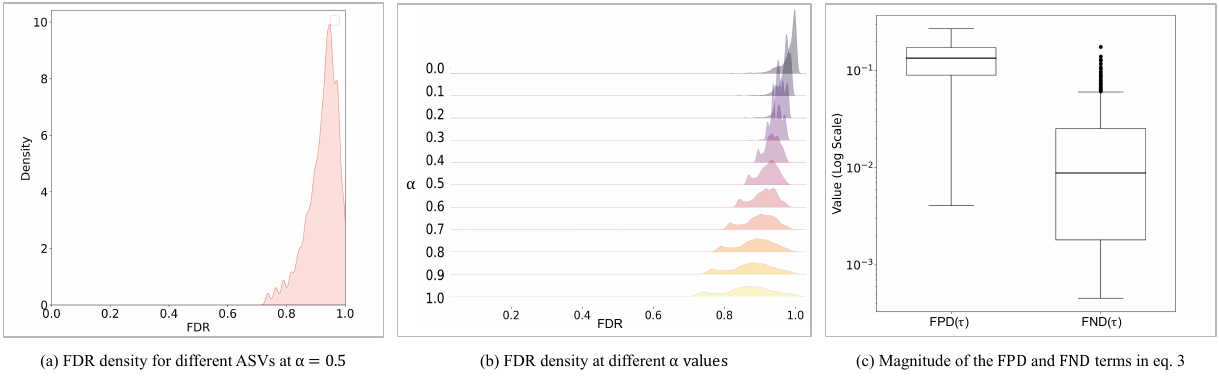}
    \caption{FDR values using 5 automatic speaker verification systems at a range of thresholds corresponding to a FMR varying form 0.1\% to 10\%}
    \label{fig:three_figures_FDR_full}
\end{figure*}
\begin{figure*}[t]
    \centering
    \includegraphics[width=1\textwidth]{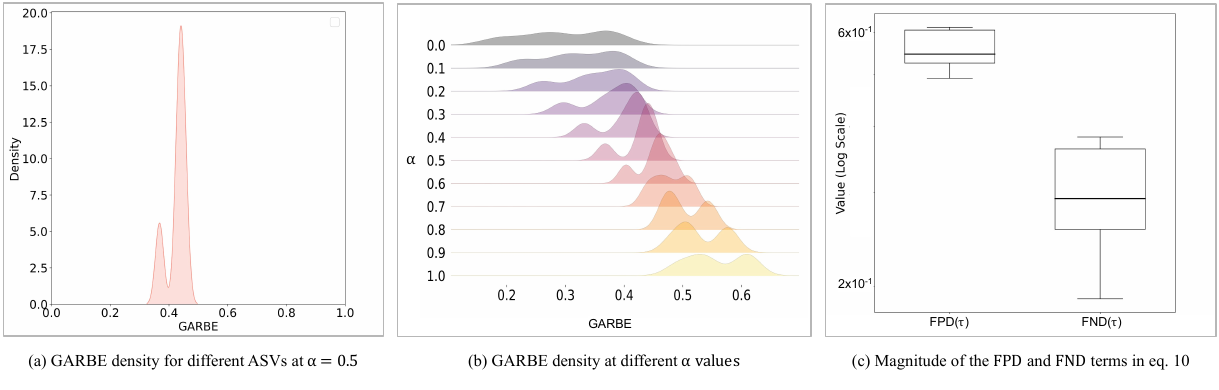}
    \caption{GARBE values using 5 automatic speaker verification systems at a threshold corresponding to FMR = 0.1\%}
    \label{fig:three_figures_GARBE_part}
\end{figure*}
\begin{figure*}[t]
    \centering
    \includegraphics[width=1\textwidth]{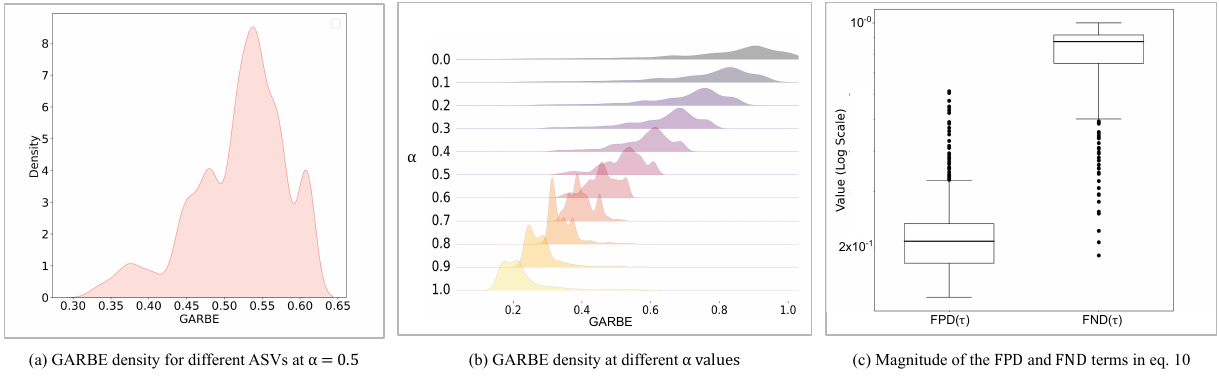}
    \caption{GARBE values using 5 automatic speaker verification systems at a range of thresholds corresponding to a FMR varying form 0.1\% to 10\%}
    \label{fig:three_figures_GARBE_full}
\end{figure*}
\begin{figure*}[t]
    \centering
    \includegraphics[width=1\textwidth]{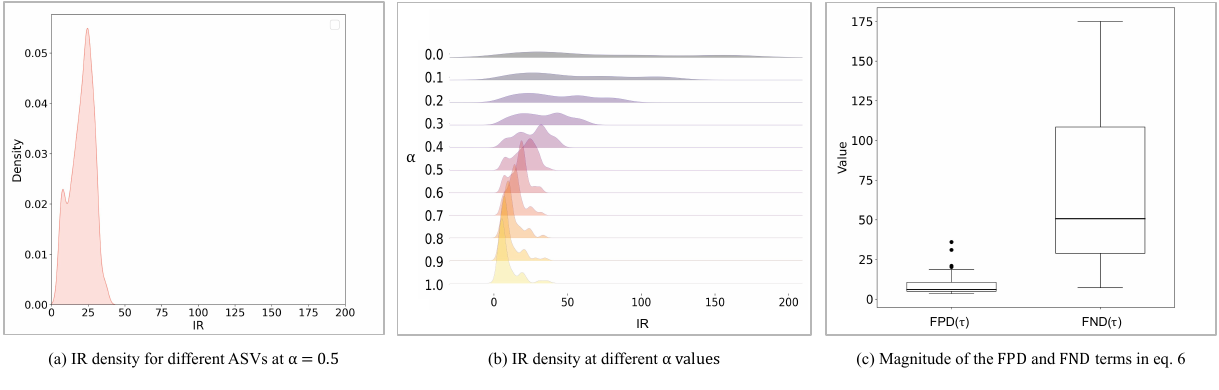}
    \caption{IR values using 5 automatic speaker verification systems at a range of thresholds corresponding to a FMR varying form 0.1\% to 10\%}
    \label{fig:three_figures_IR_full}
\end{figure*}

\subsection{Databases}
The pre-trained models\footnote{~\url{https://github.com/TaoRuijie/ECAPA-TDNN}}$^,$\footnote{~\url{https://github.com/clovaai/voxceleb_trainer}}$^,$\footnote{\url{https://github.com/alibaba-damo-academy/3D-Speaker/tree/3dspeaker}} of the ASV systems used in our experiments are trained using the development set of the VoxCeleb2 database~\cite{voxceleb2}.  It contains utterances collected from 5,994 speakers. Despite VoxCeleb2 being a multilingual dataset, the predominance of 
English speakers, and consequently also the English language, creates significant imbalance. All evaluation work is performed using the VoxCeleb1 development and test sets~\cite{voxceleb1} combined. We created a balanced protocol to assess the utility of the ASV systems. 
This protocol was created from a selection of speakers from nine different nationalities: USA, UK, Germany, Australia, Italy, India, Ireland, New Zealand, and Canada. From each of these nationality groups, eight speakers were chosen at random, resulting in a total of 72 speakers. For each speaker, 24 utterances were selected. The ASV protocol\footnote{\url{https://github.com/OubaidaOubaida/FairnessMetricsEvaluation/blob/main/pooled_data.txt}} of the pooled data consists of 39,744 comparison trials, with a balanced number of 2,208 mated and 2,208 non-mated combinations for each nationality. Due to limitations in the available data, we could not include more speakers in the study. We aimed to cover as many nationalities as possible while maintaining an equal number of speakers for each nationality. 

\subsection{Fairness evaluation procedure}

This section is dedicated to the fairness procedures and assessment of ASV systems. Presented in Table~\ref{tab:eer} are results for a preliminary analysis of ASV performance in terms of pooled EER and FMR/FNMR at the threshold corresponding to the pooled EER, across nine groups of different nationalities.
As expected, pooled EER results show that different ASV systems exhibit varying levels of performance.
The comparison also reveals consistent differences in the FMR and FNMR among nationality groups. For instance, results for the ERes2Net system are best for the UK group, 
with low FMR and FNMR. In contrast, while the same system is similarly secure for the German group (similar FMR), it does not offer the same level of convenience (higher FNMR). Interestingly, the opposite behaviour is observed for the Indian group.
Similar diverging results are observed for other nationality groups.
This analysis underscores the importance of a \emph{single} measure which reflects fairness across the \emph{full set} of groups. 
Such an approach is crucial for ensuring that ASV systems are equitable and do not disproportionately disadvantage any particular group based on nationality or other demographic factors.
 
Noting that the EER is not suited to the assessment of \emph{any} binary classifier in the case that a particular application calls for the prioritisation of a lower rate of FMR or FNMR~\cite{FDR, peri2023study}, it is similarly unsuitable as a measure of fairness.
Since the use of different operating points necessarily entails a trade-off between the FMR and the FNMR, and as argued in~\cite{FDR}, any measure of fairness must take difference in both into account. 
While one can average FMR and FNMR rates across groups, this leads to two measures which still need further interpretation if they are to provide a measure of fairness.  
This is exactly what each of the three candidate metrics delivers. 

Our methodology to evaluate the proposed fairness metrics is aligned with that in~\cite{face_metrics_eval}.
To this end, we adopt a benchmark threshold that corresponds to an FMR of 0.1\% for the initial evaluation. 
In addition, and going beyond the work in~\cite{face_metrics_eval}, we follow the requirements of ISO/IEC DIS 19795-10~\cite{ISO-IEC-DIS-19795-10-230928} 
by investigating a range of thresholds, specifically within an FMR interval of 0.1\%-10\%, alongside a comprehensive risk parameter range of [0,1]. The rationale behind the selection of a 0.1\%-10\% FMR interval is twofold: first, though it depends on the application, today's most advanced ASV systems achieve acceptable levels of FNMR at FMRs in the order of 0.1\%; second, ASV systems (or indeed any other biometric system) with an FMR surpassing 10\% might not limited use. Our objective is to examine any variation in fairness at a range of representative operating points, for five distinct systems, thereby gaining a holistic understanding of metric behaviour. This examination is crucial as, in real-world scenarios, each system operates optimally at a threshold that is contingent upon the specific application it serves. 

\section{Experimental results and discussion}
In this section, we present an assessment of the fairness metrics presented in Section~\ref{sec:metrics}, first for a fixed threshold which provides an FMR = 0.1\% and, second, for a range of thresholds from 0.1\% to 10\%. Last, we evaluate each fairness metric in terms of the three FFMCs described in Section~\ref{sec:FFMCs}.

\subsection{Metrics evaluation results at a fixed threshold}
\label{sec:eval_fixed_tau}
We evaluate the fairness metrics  performance across five different ASV systems. We aggregate the evaluation results to understand how these metrics behave under different systems. 

For each metric we present three plots. The first is the metric density distribution when the importance of the FMR/FNMR differentials are equal ($\alpha$ =0.5), such as presented in the reference work~\cite{face_metrics_eval}. The second is to show the density distribution of the metric for all alpha values in [0,1]. These two plots show the spread of the metrics values to ascertain whether the metric offers an intuitive comparison between the systems, as well as their behavior under varying risk parameter alpha. The third graph is the distribution of the \text{FPD} and \text{FND} terms for alpha in [0,1] to study the scales of the error rates. 

To align our work with that in~\cite{face_metrics_eval}, we evaluated the fairness metric at a fixed threshold.
Experiments were performed with decision thresholds $\tau$ set so as to produce a FMR of 0.1\%

\subsubsection{FDR evaluation}
Our assessment starts with the calculation of the FDR metric described in Section~\ref{sec:fdr}.
The FDR values are visually depicted in Figure~\ref{fig:three_figures_FDR_part}. 
Figure~\ref{fig:three_figures_FDR_part}(a) and~\ref{fig:three_figures_FDR_part}(b) show FDR values concentrated in the range [0.82-1]. It is not intuitive to compare which system is more fair and to assess the impact of the risk parameter $\alpha$, especially in the case of mostly fair systems. Figure~\ref{fig:three_figures_FDR_part}(c) illustrates that the differential terms (\text{FPD}, related to the FMR, and \text{FND}), related to the FNMR, operate on significantly disparate scales. Aggregating these terms presents a challenge in accurately configuring them with the term $\alpha$. This complexity underscores the difficulty of intuitively measuring the contributions of FMR and FNMR within the FDR metric. Consequently, this approach fails to satisfy the criteria established by the first FFCM principle (Section~\ref{sec:FFMCs}).

\subsubsection{IR evaluation}
The evaluation of the IR metric presented in Section~\ref{sec:ir} revealed minFMR values of 0 for certain subgroups. This  renders the computation of the \text{FPD} term of the IR equation~(\ref{eq:A_IR}) unfeasible, subsequently leading to the inability to calculate the IR. Among the five evaluated ASV systems, the IR metric is only computable for the ResNetSEV2 system, with the value being 13.35. This shows that the IR metric does not fulfill the third FFCM (Section~\ref{sec:FFMCs}).
Consequently, this casts doubt on the suitability of IR as a reliable metric for fairness assessment in such contexts. Additionally, the ratio nature of the IR introduces another complexity: there is no upper limit to the IR values. This means that the IR can vary greatly and potentially reach extremely high numbers, further complicating its interpretation.

\subsubsection{GARBE evaluation}
We now turn our attention to the GARBE metric described in Section~\ref{sec:garbe}. The results illustrated in Figure~\ref{fig:three_figures_GARBE_part}(a) and~\ref{fig:three_figures_GARBE_part}(b) exhibit a wider range compared to the FDR values. This range extends from 0.19 to 0.61 for $\alpha$ in [0,1], which is about half of the theoretical range. This provides a more intuitive comparison between the systems and assessment of the impact of $\alpha$. An additional critical finding is related to the \text{FPD} and \text{FND} terms, depicted in Figure~\ref{fig:three_figures_GARBE_part}(c). These terms are scaled to a comparable magnitude. Specifically, the median value for the \text{FPD} term is identified at 0.55, while the median for the \text{FND} term is observed at 0.29. The normalised Gini coefficient computation reduced the discrepancy of the differential terms scale. The impact of $\alpha$ is then more pronounced. 

\subsection{Metrics evaluation results at different thresholds}
\label{sec:eval_all_tau}
In an extension of the previous work, we expand the assessment to include not just different systems, but also different systems with different operating points. We expanded the range of our analysis by varying the threshold of the five ASV systems from that corresponding to a 0.1\% to 10\% FMR. This approach ensures an assessment reflective of real-world scenarios.

For each metric we present three plots as in the previous section. The only difference this time is the FMR of the plots range from 0.1\% to 10\%.

\subsubsection{FDR evaluation}
Despite adjusting both the threshold and $\alpha$ values, the FDR values display a consistent trend, with values concentrated between 0.72 and 1, as shown in Figures~\ref{fig:three_figures_FDR_full}(a) and~\ref{fig:three_figures_FDR_full}(b). It is still not intuitive to compare which system is more fair as most of the FDR values almost overlap for all \textit{alphas}.  Moreover, the scale disparity between the \text{FPD} and \text{FND} terms is still persistent, as depicted in Figure~\ref{fig:three_figures_FDR_full}(c). This implies the intuitiveness of the contributions of FMR and FNMR to the FDR metric. Hence, the FDR metric does not meet the FFMC.1.

\subsubsection{IR evaluation}
Results for the IR metric confirm the impossibility of its computation in certain scenarios, even with a variable threshold range. The FFMC.3 is still not fulfilled. The selective representation of 8.6\% of computable values in Figures~\ref{fig:three_figures_IR_full}(a) and~\ref{fig:three_figures_IR_full}(b) illustrates the extensive range of the IR, reaching up to 200 in certain instances. This confirms that the IR metric is not bounded, therefore, it does not meet FMMC.2. However,the \text{FPD} and \text{FND} terms are on a similar scale, as demonstrated in Figure~\ref{fig:three_figures_IR_full}(c). Use of a ratio-based approach means that the IR metric offers a more balanced comparison between terms. Thus, the FFMC.1. is fulfilled.

\subsubsection{GARBE evaluation}
The GARBE metric is effective in avoiding the limitations seen in the previous metrics effectively. As Figure~\ref{fig:three_figures_GARBE_full}(a) and~\ref{fig:three_figures_GARBE_full}(b) show, GARBE values consistently span the full theoretical range from 0 to 1. Furthermore, Figure~\ref{fig:three_figures_GARBE_full}(b) demonstrates the sensitivity of GARBE to \textit{alpha} changes. The \text{FPD} and \text{FND} terms, presented in Figure~\ref{fig:three_figures_GARBE_part}(c), with median values of 0.21 and 0.88 respectively, are on the same scale. This ensures that both terms contribute significantly to the fairness assessment, meeting FFMC.1.

Analysis of the boxplots shows a swap in the boxes of the \text{FPD} and \text{FND} terms between Figures~\ref{fig:three_figures_FDR_part}(c) and ~\ref{fig:three_figures_FDR_full}(c), as well as between ~\ref{fig:three_figures_GARBE_part}(c) and~\ref{fig:three_figures_GARBE_full}(c). The swap occurs because the \text{FPD} is higher or lower than the \text{FND} before reaching certain thresholds, and vice versa. More precisely, within the FDR metric analysis, for thresholds that yield an FMR lower than 0.9\%, the \text{FND} term surpasses the \text{FPD} term. Conversely, for thresholds resulting in an FMR above 0.9\%, the \text{FPD} term is higher. Regarding the GARBE metric, thresholds that produce an FMR below 0.4\% result in a higher \text{FPD} than \text{FND}. This explains the observed variation in the boxplot positions when the FMR is set at 0.1\% and for FMR ranges from 0.1\% to 10\%.

\subsection{Summary of the Fairness Metrics Criteria}
Our evaluation of fairness metrics for ASV within the framework of the Functional Fairness Measure Criteria (Section~\ref{sec:FFMCs}) presents varied results for the FDR, IR, and GARBE metrics. These outcomes are  summarised in Table~\ref{tab:summ}.

The FDR metric is inherently bounded therefore satisfies the FFMC.2 criterion. This quality provides an interpretable measure of fairness with a value of 1 to a perfectly fair system and a value of 0 to a perfectly unfair system. The FDR metric also maintains computability when FNMR or FMR is zero, adhering to FFMC.3. However, the differential terms \text{FPD} and \text{FND} exist at vastly different scales when using a normal range of risk parameters ($\alpha$ in [0,1]) and operationally relevant error rates (\text{FMR} in [0.1\%,10\%]). This renders the interpretation of the contribution of the FMR and FNMR in the computation of the FDR metric challenging. Hence, this metric does not meet FFMC.1.

The IR metric fulfills FFMC.1 by taking a ratio approach which better balances the contributions of \text{FPD} and \text{FND} terms. Nevertheless, it faces limitations in meeting FFMC.2 and FFMC.3 due to its unbounded nature, which makes it challenging to set benchmarks and renders it incalculable when FNMR or FMR reach zero.

GARBE emerges as the most robust metric, fully meeting all FFMC criteria. By employing the Gini coefficient, as specified in Equations~\ref{eq:Gb_FMR} and~\ref{eq:Gb_FNMR}, the \text{FPD} and \text{FND} terms are converted to a same scale before their aggregation.  
This normalisation is key to fulfilling FFMC.1, as it makes the contribution of the FMR and FNMR to the calculation of the GARBE metric intuitive. It ensures a nuanced and balanced representation of the metric measures across varying $\alpha$ values. 
Second, GARBE meets FFMC.2 by maintaining set boundaries, which allows for the establishment of clear benchmarks. Furthermore, it remains calculable even where error rates are zero, thereby meeting FFMC.3. 

Our analysis validates and extends the findings of the study of fairness for face recognition reported in~\cite{face_metrics_eval}, reinforcing the conclusion that the GARBE metric stands out as the most suitable for evaluating fairness in biometric systems. This congruence illustrates that GARBE are not confined to face recognition systems but extend also to studies of fairness in automatic speaker verification. 
\begin{table}[!t]
\centering
\caption{Summary of Fairness Measures Criteria for ASV}
\begin{tabular}{|l|c|c|c|}
\hline
FFMC Criteria & FDR & IR & GARBE \\ \hline
FFMC.1        &  &  \checkmark   & \checkmark \\ \hline
FFMC.2        & \checkmark &     & \checkmark \\ \hline
FFMC.3        & \checkmark &     & \checkmark \\ \hline
\end{tabular}
\label{tab:summ}
\end{table}

\begin{figure}[!t]
\centering
\includegraphics[width=\linewidth]{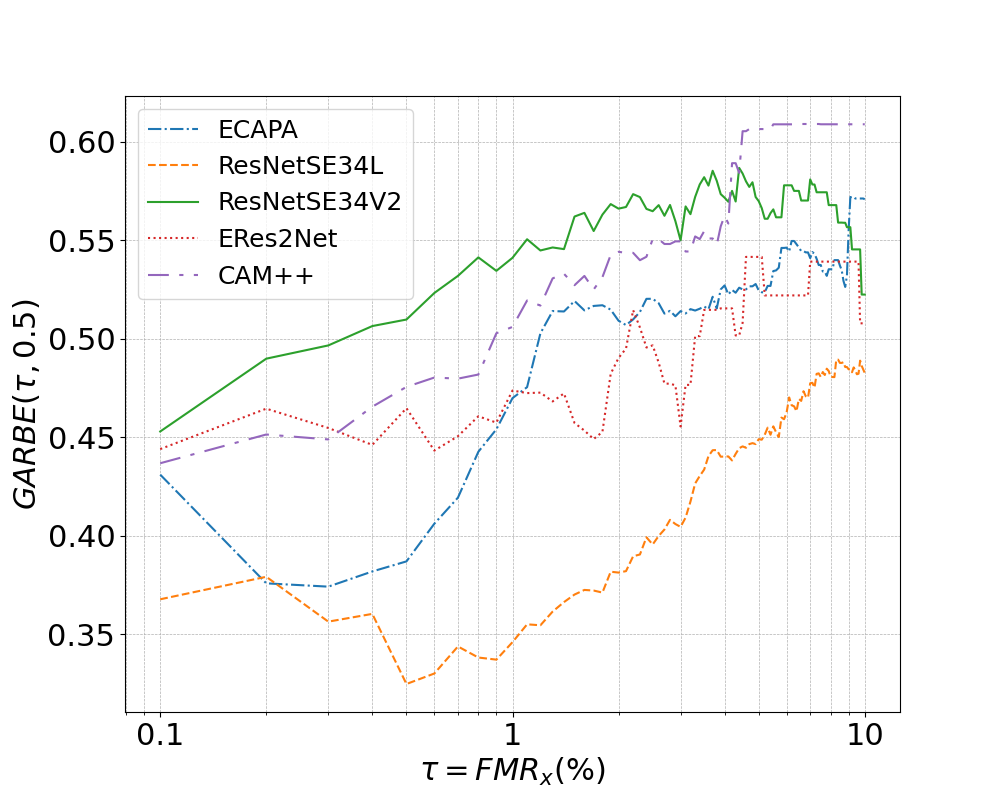}
\caption{GARBE of different ASV systems for different decision $\tau = \text{FMR}_x$ where $x$ varies from 0.1\% to 10\% and for $\alpha = 0.5$.}
\label{fig:garbe_systems}
\end{figure}

\begin{figure}[!h]
\centering
\includegraphics[width=\linewidth]{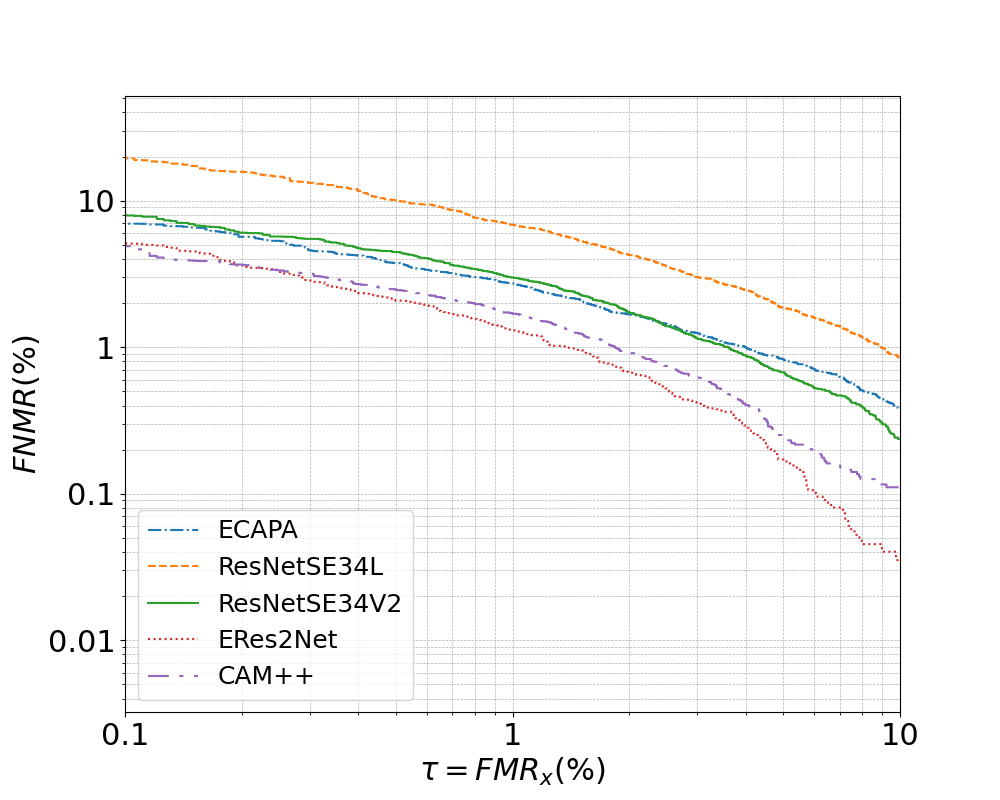}
\caption{Detection error tradeoff (DET) curve of different ASV systems.}
\label{fig:det_curve_systems}
\end{figure}

\section{Fairness and ASV assessment}
While it is not the primary objective of this work, our attention now shifts to an analysis of the ASV systems themselves. 
We report an assessment of the five ASV systems in terms of verification performance (FMR vs.\ FNMR) and fairness, here assessed using only the GARBE metric. 

Figure~\ref{fig:garbe_systems} shows a plot of GARBE values over an FMR range of between 0.1\% and 10\%, for $\alpha = 0.5$.  
Figure~\ref{fig:det_curve_systems} shows a corresponding detection error trade-off (DET) plot.
The ResNetSE34L system shows the lowest GARBE values across all FMR thresholds, indicating that the system is the fairest.
However, in terms of verification performance, the same ResNetSE34L system is the worst. 
Improved fairness (lower differential performance  across groups) in this case comes at the cost of degraded verification performance. 

GARBE values for the least fair system, ResNetSE34V2, are highest across a majority of operating points, indicating significant differential performance across groups.
Verification accuracy is, however, better than that for the ResNetSE34L system, another indication of an apparent trade-off between performance and fairness.
The ERes2Net and CAM++ systems are not only comparably fair, albeit not \emph{the} fairest, especially for FMRs under 1\%, they are among the best performing.

The ECAPA model exhibits different behaviour. Verification performance is average, but there is substantial variation in GARBE values between lower and higher FMRs, more so than for other systems.
This pronounced fluctuation suggests a particular sensitivity for the ECAPA model, 
a distinct aspect of its operational characteristics.

\section{Conclusions}
We present a comparison of three different fairness metrics — FDR, IR, and GARBE — in the context of automatic speaker verification (ASV) 
alongside an analysis of fairness and verification performance for five \textit{state-of-the-art} ASV systems. Results show that the Gini Aggregation Rate for Biometric Equitability (GARBE) metric stands out as that which best fulfills the Functional Fairness Measure Criteria (FFMCs).

Our analysis reveals a delicate balance between fairness and accuracy. The fairest system has the worst verification performance.  The system with the best verification performance has average fairness.  These findings show the challenge to strike a balance between fairness and verification performance.
Given the requirement for fairness, it should be evaluated during the development of ASV systems, as it should be during the development of any biometric system; there is no guarantee that a focus on raw verification performance alone will produce equitable solutions. We recommend as future work to develop \emph{fairness by design} concepts.

\section{Acknowledgements}
This work is supported by the TReSPAsS-ETN project funded by the European Union’s Horizon 2020 research and innovation programme under the Marie Skłodowska-Curie grant agreement No. 860813.

\bibliographystyle{IEEEbib}
\bibliography{mybib.bib}

\begin{thebibliography}{10}

\bibitem{ISO-IEC-DIS-19795-10-230928}
{ISO/IEC JTC1 SC37 Biometrics},
\newblock {\em {ISO/IEC} {DIS} 19795-10. Information Technology -- Biometric Performance Testing and Reporting -- Part~10: Quantifying biometric system performance variation across demographic groups},
\newblock International Organization for Standardization, 2023.

\bibitem{ISO-IEC-19795-1-Framework-210216}
{ISO/IEC JTC1 SC37 Biometrics},
\newblock {\em {ISO/IEC} 19795-1:2021. Information Technology -- Biometric Performance Testing and Reporting -- Part~1: Principles and Framework},
\newblock International Organization for Standardization, June 2021.

\bibitem{Drozdowski-BiasSurvey-TTS-2020}
Pawel Drozdowski, Christian Rathgeb, Antitza Dantcheva, Naser Damer, and Christoph Busch,
\newblock ``Demographic bias in biometrics: A survey on an emerging challenge,''
\newblock {\em Trans. on Technology and Society ({TTS})}, vol. 1, no. 2, pp. 89--103, June 2020.

\bibitem{Drozdowski-WatchlistImbalanceEffect-ICCVW-2021}
Pawel Drozdowski, Christian Rathgeb, and Christoph Busch,
\newblock ``The watchlist imbalance effect in biometric face identification: Comparing theoretical estimates and empiric measurements,''
\newblock in {\em Intl. Conf. on Computer Vision Workshops ({ICCVW})}, New York, October 2021, pp. 1--9, IEEE.

\bibitem{garbe}
John Howard, Eli~Laird andYevgeniy Sirotin, Rebecca Rubin, Jerry Tipton, and Arun Vemury,
\newblock ``Evaluating proposed fairness models for face recognition algorithms,''
\newblock in {\em Proc. Intl. Conf. on Pattern Recognition}, 2022.

\bibitem{Rathgeb-FairnessExperts-TSM-2022}
Christian Rathgeb, Pawel Drozdowski, Dinusha~C. Frings, Naser Damer, and Christoph Busch,
\newblock ``Demographic fairness in biometric systems: What do the experts say?,''
\newblock {\em {IEEE} Technology and Society Magazine}, vol. 41, pp. 71--82, December 2022.

\bibitem{Kotwal-FairnessIndex-ICPR-2022}
Ketan Kotwal and Sébastien Marcel,
\newblock ``Fairness index measures to evaluate bias in biometric recognition,''
\newblock in {\em Proc. Intl. Conf. on Pattern Recognition}, 2022.

\bibitem{FDR}
Tiago de~Freitas~Pereira and Sébastien Marcel,
\newblock ``Fairness in biometrics: A figure of merit to assess biometric verification systems,''
\newblock {\em IEEE Transactions on Biometrics, Behavior, and Identity Science}, vol. 4, no. 1, pp. 19--29, 2022.

\bibitem{estevez2022study}
Mariel Estevez and Luciana Ferrer,
\newblock ``Study on the fairness of speaker verification systems on underrepresented accents in english,''
\newblock {\em arXiv preprint arXiv:2204.12649}, 2022.

\bibitem{peri2023study}
Raghuveer Peri, Krishna Somandepalli, and Shrikanth Narayanan,
\newblock ``A study of bias mitigation strategies for speaker recognition,''
\newblock {\em Computer Speech \& Language}, vol. 79, pp. 101481, 2023.

\bibitem{chouchane2023fairness}
Ouba{\"\i}da Chouchane, Michele Panariello, Chiara Galdi, Massimiliano Todisco, and Nicholas Evans,
\newblock ``Fairness and privacy in voice biometrics: A study of gender influences using wav2vec 2.0,''
\newblock in {\em 2023 International Conference of the Biometrics Special Interest Group (BIOSIG)}. IEEE, 2023, pp. 1--7.

\bibitem{deltas2003small}
George Deltas,
\newblock ``The small-sample bias of the gini coefficient: results and implications for empirical research,''
\newblock {\em Review of economics and statistics}, vol. 85, no. 1, pp. 226--234, 2003.

\bibitem{face_metrics_eval}
John~J. Howard, Eli~J. Laird, Rebecca~E. Rubin, Yevgeniy~B. Sirotin, Jerry~L. Tipton, and Arun~R. Vemury,
\newblock ``Evaluating proposed fairness models for face recognition algorithms,''
\newblock in {\em International Conference on Pattern Recognition}. Springer, 2022, pp. 431--447.

\bibitem{ecapa_repo}
Rohan~Kumar Das, Ruijie Tao, and Haizhou Li,
\newblock ``Hlt-nus submission for 2020 nist conversational telephone speech sre,''
\newblock {\em arXiv preprint arXiv:2111.06671}, 2021.

\bibitem{ecapa}
Brecht Desplanques, Jenthe Thienpondt, and Kris Demuynck,
\newblock ``{ECAPA-TDNN}: Emphasized channel attention, propagation and aggregation in tdnn based speaker verification,''
\newblock in {\em Proc. INTERSPEECH 2020}, 2020, pp. 3830--3834.

\bibitem{ResNetSE34L}
Joon~Son Chung, Jaesung Huh, Seongkyu Mun, Minjae Lee, Hee~Soo Heo, Soyeon Choe, Chiheon Ham, Sunghwan Jung, Bong-Jin Lee, and Icksang Han,
\newblock ``In defence of metric learning for speaker recognition,''
\newblock in {\em Proc. INTERSPEECH 2020}, 2020, pp. 2977--2981.

\bibitem{resnet}
Kaiming He, Xiangyu Zhang, Shaoqing Ren, and Jian Sun,
\newblock ``Deep residual learning for image recognition,''
\newblock in {\em Proceedings of the IEEE conference on computer vision and pattern recognition}, 2016, pp. 770--778.

\bibitem{SAP}
Weicheng Cai, Jinkun Chen, and Ming Li,
\newblock ``Exploring the encoding layer and loss function in end-to-end speaker and language recognition system,''
\newblock in {\em Proc. The Speaker and Language Recognition Workshop (Odyssey 2018)}, 2018, pp. 74--81.

\bibitem{ResNetSE34V2}
Yoohwan Kwon, Hee-Soo Heo, Bong-Jin Lee, and Joon~Son Chung,
\newblock ``The ins and outs of speaker recognition: lessons from voxsrc 2020,''
\newblock in {\em ICASSP 2021-2021 IEEE International Conference on Acoustics, Speech and Signal Processing (ICASSP)}. IEEE, 2021, pp. 5809--5813.

\bibitem{ASP}
Koji Okabe, Takafumi Koshinaka, and Koichi Shinoda,
\newblock ``Attentive statistics pooling for deep speaker embedding,''
\newblock in {\em Proc. INTERSPEECH 2018}, 2018, pp. 2252--2256.

\bibitem{ERes2Net}
Yafeng Chen, Siqi Zheng, Hui Wang, Luyao Cheng, Qian Chen, and Jiajun Qi,
\newblock ``{An enhanced {R}es2{N}et with local and global feature fusion for speaker verification},''
\newblock in {\em Proc. INTERSPEECH 2023}, 2023, pp. 2228--2232.

\bibitem{cam++}
Hui Wang, Siqi Zheng, Yafeng Chen, Luyao Cheng, and Qian Chen,
\newblock ``Cam++: A fast and efficient network for speaker verification using context-aware masking,''
\newblock in {\em Proc. INTERSPEECH 2023}, 2023, pp. 5301--5305.

\bibitem{voxceleb2}
Joon~Son Chung, Arsha Nagrani, and Andrew Zisserman,
\newblock ``Voxceleb2: Deep speaker recognition,''
\newblock in {\em Proc. INTERSPEECH 2018}, 2018, pp. 1086--1090.

\bibitem{voxceleb1}
Arsha Nagrani, Joon~Son Chung, and Andrew Zisserman,
\newblock ``Voxceleb: a large-scale speaker identification dataset,''
\newblock in {\em Proc. INTERSPEECH 2017}, 2017, pp. 2616--2620.

\end{thebibliography}

\end{document}